\documentclass[twocolumn,superscriptaddress,showpacs,preprintnumbers,amsmath
,amssymb,aps,prd]{revtex4}
\usepackage{graphicx}
\usepackage{bm}
\usepackage{color}

\def\ltsima{$\; \buildrel < \over \sim \;$}
\def\ltsim{\lower.5ex\hbox{\ltsima}}
\def\gtsima{$\; \buildrel > \over \sim \;$}
\def\gtsim{\lower.5ex\hbox{\gtsima}}

\newcommand{\KMS}{\mbox{km s}^{-1}\,}

\def\newacronym#1#2#3{\gdef#1{#3 (#2)\gdef#1{#2}}}

\def\apjl{Astrophys. J. Lett.}
\def\mnras{MNRAS} 

\newacronym{\NSF}{NSF}{National Science Foundation}
\newacronym{\NASA}{NASA}{National Aeronautics and Space Administration}
\newacronym{\lisa}{LISA}{the Laser Interferometer Space Antenna}
\newacronym{\ligo}{LIGO}{Laser Interferometer Gravitational-wave Observatory} 
\newacronym{\Caltech}{Caltech}{California Institute of Technology}
\newacronym{\MIT}{MIT}{Massachusetts Institute of Technology}
\newacronym{\sph}{SPH}{smooth particle hydrodynamics}
\newacronym{\tsi}{TSI}{the Terascale Supernova Initiative}
\newacronym{\wmap}{WMAP}{the Wilkinson Microwave Anisotropy Probe}
\newacronym{\decigo}{DECIGO}{the Deci-Hertz Interferometric Gravitational-wave Observatory} 
\newacronym{\cmbr}{CMBR}{cosmic microwave background}
\newacronym{\ibbh}{IBBH}{intermediate binary black hole}
\newacronym{\bdj}{BDJ}{Brans-Dicke-Jordan}
\newacronym{\bbo}{BBO}{Big Bang Observer}
\newacronym{\decigo}{DECIGO}{Deci-Hertz Gravitational-Wave Observatory}

\def\MPR#1{{\it Moving Puncture Recipe}#1 (MPR#1)\gdef\MPR{MPR}}
\def\ahz#1{apparent horizon#1 (AH#1)\gdef\ahz{AH}}
\def\CLA#1{close-limit approximation#1 (CLA#1)\gdef\CLA{CLA}}
\def\NR#1{numerical relativity#1 (NR#1)\gdef\NR{NR}}
\def\pnw#1{post-Newtonian#1 (PN#1)\gdef\pnw{PN}}
\def\qnm#1{quasi-normal mode#1 (QNM#1)\gdef\qnm{QNM}}
\def\isco#1{innermost stable circular orbit#1 (ISCO#1)\gdef\isco{ISCO}}
\def\eos#1{equation of state#1 (EOS#1)\gdef\eos{EOS}}
\def\tov#1{Tolman-Oppenheimer-Volkoff#1 (TOV#1)\gdef\tov{TOV}}
\def\ns#1{neutron star#1 (NS#1)\gdef\ns{NS}}
\def\bbh#1{binary black holes#1 (BBH#1)\gdef\bbh{BBH}}
\def\bhns#1{black hole -- neutron star#1 (BHNS#1)\gdef\bhns{BHNS}}
\def\nsns#1{neutron star -- neutron star#1 (NSNS#1)\gdef\nsns{NSNS}}
\def\emri#1{extreme mass-ratio inspiral#1 (EMRI#1)\gdef\emri{EMRI}}
\def\emrb#1{extreme mass-ratio binaries#1 (EMRB#1)\gdef\emrb{EMRB}} 
\def\grb#1{gamma-ray burst#1 (GRB#1)\gdef\grb{GRB}}
\def\imbh#1{intermediate mass black hole#1 (IMBH#1)\gdef\imbh{IMBH}}
\def\smbh#1{supermassive black hole#1 (SMBH#1)\gdef\smbh{SMBH}}
\def\bh#1{black hole#1 (BH#1)\gdef\bh{BH}}
\def\ulx#1{ultra-luminous x-ray source#1 (ULX#1)\gdef\ulx{ULX}}
\def\lmxbs{low-mass x-ray Binaries (LMXBs)\gdef\lmxbs{LMXBs}\gdef\lmxb{LMXB}} 
\def\lmxb{low-mass x-ray Binary (LMXB)\gdef\lmxbs{LMXBs}\gdef\lmxb{LMXB}} 

\def\kickmax{$10,000\, \mathrm{km \, s^{-1}}$}

\begin{document}

\title{Superkicks in Hyperbolic Encounters of Binary Black Holes}

\author{James Healy}
\affiliation{Center for Gravitational Wave Physics\\
The Pennsylvania State University, University Park, PA 16802}
\author{Frank Herrmann}
\affiliation{Center for Gravitational Wave Physics\\
The Pennsylvania State University, University Park, PA 16802}
\author{Ian Hinder}
\affiliation{Center for Gravitational Wave Physics\\
The Pennsylvania State University, University Park, PA 16802}
\author{Deirdre M. Shoemaker}
\affiliation{Center for Gravitational Wave Physics\\
The Pennsylvania State University, University Park, PA 16802}
\affiliation{Center for Relativistic Astrophysics and
School of Physics\\
Georgia Institute of Technology, Atlanta, GA 30332}
\author{Pablo Laguna}
\affiliation{Center for Gravitational Wave Physics\\
The Pennsylvania State University, University Park, PA 16802}
\affiliation{Center for Relativistic Astrophysics and
School of Physics\\
Georgia Institute of Technology, Atlanta, GA 30332}
\author{Richard A. Matzner}
\affiliation{Center for Relativity and Department of Physics\\
The University of Texas at Austin, Austin, TX 78712}

\begin{abstract}
  Generic inspirals and mergers of binary black holes produce beamed
  emission of gravitational radiation that can lead to a
  gravitational recoil or {\em kick} of the final black hole.  The kick
  velocity depends on the mass ratio and spins of the binary as well
  as on the dynamics of the binary configuration. Studies have focused
  so far on the most astrophysically relevant configuration of
  quasi-circular inspirals, for which kicks as large as $\sim 3,300\,
  \KMS$ have been found. We present the first study of
  gravitational recoil in {\em hyperbolic} encounters. Contrary to
  quasi-circular configurations, in which the beamed radiation tends
  to average during the inspiral, radiation from hyperbolic encounters
  is plunge dominated, resulting in an enhancement of preferential
  beaming. As a consequence, it is possible to achieve kick velocities
  as large as \kickmax.
\end{abstract}


\maketitle 

Numerical relativity estimates of the gravitational recoil or kick
inflicted on the final \bh{} from generic inspirals and mergers of
\bbh{} have triggered tremendous excitement in astrophysics. This is
mainly due to the fact that most galaxies host a \smbh{} at their
centers~\cite{1998Natur.395A..14R,2007MNRAS.381..136D}. As galaxies
merge, a kick to the final \bh{} from the coalescence of the \bh{s} at
the galactic cores could have profound implications in subsequent
mergers, affecting the growth of \smbh{s} via mergers as well as the
population of galaxies containing \smbh{s}. In addition, there have
been several suggestions of direct observational signatures of
putative \bh{} recoils~\cite{2008arXiv0802.3873S,2008arXiv0805.2609D,
  2008arXiv0805.1420B,2008ApJ...676L...5L,
  2007PhRvL..99d1103L,2008arXiv0803.0003K}, with one
study~\cite{2008ApJ...678L..81K} presenting evidence for the first
candidate of a recoiling \smbh{.}

\bh{} kick velocities depend on the mass ratio and spins of the
merging \bh{s} as well as on the initial configuration and subsequent
dynamics of the binary. Studies published to date have concerned the
most astrophysically relevant configuration, that of quasi-circular
inspirals~\cite{2007CQGra..24...33H,2007PhRvL..98i1101G,
  2007ApJ...661..430H,2007PhRvL..99d1102K}.  One remarkable discovery
has been kicks of a few thousand $\KMS$ found in configurations of
equal-mass binaries with initially anti-aligned spins in the orbital
plane~\cite{Gonzalez:2007hi,
  2007ApJ...659L...5C,2007PhRvL..98w1102C}. For near extremal spins
($a/m = 0.925$), recoils as large as $3,300\, \KMS$ have been
computed~\cite{2008arXiv0803.0351D}.

Motivated by our previous study~\cite{2008arXiv0802.2520W} of the
final spin of \bh{s} from scattering mergers of \bbh{s}, we present
the first extension of gravitational recoil to hyperbolic
encounters. There are crucial differences between hyperbolic and
quasi-circular configurations that affect the kick to the final \bh{.}
For quasi-circular orbits, the emitted radiation is asymmetrically
beamed, and carries linear momentum with it. But it tends to average
out during the inspiral~\cite{2005ApJ...635..508B}, producing a modest
wobbling and drift of the center of mass of the binary. The final kick
arises because as the binary approaches the plunge, the
\emph{averaging} loses its effectiveness, leading to a gradual recoil
build-up. Both numerical simulations and post-Newtonian
studies~\cite{2005ApJ...635..508B,2006PhRvD..73l4006D} have confirmed
the gradual kick accumulation during the inspiral, and, in addition,
the studies have shown that most of the recoil is generated during the
plunge. In some instances, during the plunge and ring-down there is
also a period of \emph{anti-kick} before reaching the final kick
value~\cite{2007ApJ...668.1140B,2008PhRvD..77d4031S}.

The main motivation for the present study was to consider
plunge-dominated configurations to investigate whether kicks
comparable to those for quasi-circular inspirals can be found. We were
surprised to find that kicks larger than $3000\, \KMS$ are in general
produced for spin configurations equivalent to those studied in
quasi-circular inspirals. We show below a configuration giving kick
velocities as large as \kickmax{.}  Two qualitative features of
hyperbolic encounters contribute to these larger kicks.  Not only are
hyperbolic encounters plunge dominated, but the nature of the plunge
is such that it enhances the beaming of radiated linear
momentum~\cite{Miller:2008en}.

We use the same computational infrastructure and methodology as in
previous studies~\cite{2008arXiv0802.2520W,
  2007ApJ...661..430H,2007CQGra..24...33H}, namely a BSSN code with
moving puncture gauge conditions.  The hyperbolic encounters are
initiated with Bowen-York initial data~\cite{Bowen:1980yu}. The data
consist of two equal-mass \bh{s} with masses $m=M/2$ located along the
$x$-axis: \bh{$_\pm$} is located at $x = \pm 5\,M$ and has linear
momentum $\vec{P}_\pm = \pm (P\,\cos{\theta}, P\,\sin{\theta},0)$ with
$\theta$ the angle in the orbital plane with respect to the
$x$-axis. The total initial orbital angular momentum is then
$\vec{L}/M^2 = 10\,(P/M)\,\sin{\theta}\,\hat z$. The spins of each
\bh{} are in the orbital plane: \bh{$_\pm$} has spin $\vec{S}_\pm =
\pm (S\,\cos{\phi_\pm},S\,\sin{\phi_\pm},0)$, with $\phi_\pm$ the
angle in the orbital plane with respect to the $\pm x$-axis. The
parameter space of our encounters is quite large: $\lbrace P, \theta,
S, \phi_\pm\rbrace$. However, from exploratory runs we have gained a
good understanding of the parameter space and isolated those
parameters that can be kept fixed without seriously compromising the
goals of the study.

For most of the runs, we have kept the spin magnitudes at $S/M^2 =
0.2$ ($a/m = 0.8$), with the exception of those runs used to
investigate the dependence of the kick on $S$.  We kept fixed also the
impact angle at $\theta=153.4^{\circ}$. We considered some other angles
but found that this is the angle for which we obtained the largest
kicks.  Finally, for most cases we kept the spin direction of
\bh{$_-$} located at $x= -5\,M$ fixed at $\phi_- = 0^\circ$ or
$45^\circ$. Once other parameters were fixed from among a small set of
values, the parameter we varied in general was the linear momentum
magnitude $P$.

Kicks are computed from a surface integral~\cite{1980grg2.conf....1N,Campanelli:1998jv}
involving the Weyl curvature tensor $\Psi_4$. Although strictly
speaking this kick formula must be evaluated in the limit $r
\rightarrow \infty$, we applied it at extraction radii $r/M = \lbrace
40,50,75,85,100 \rbrace$. The resultant kicks were fitted to both
$V(r) = V_\infty^{(1)} + K_0/r$ and $V(r) = V_\infty^{(2)} + K_1/r+
K_2/r^2$. The extrapolated $r \rightarrow \infty$ kicks and their
errors were estimated from $V_\infty = (V_\infty^{(1)} +
V_\infty^{(2)})/2$ and $\delta V_\infty = |V_\infty^{(1)} -
V_\infty^{(2)}|$, respectively. Unless explicitly noted, all the
reported kick velocities and energy radiated were obtained from
simulations with resolution  $h = M/0.8$ on
the mesh used for the kick computation.
Every run had 10 levels of factor-of-2
refinement, with outer boundaries at $\sim 320\,M$. We discuss below the
convergence of kick estimates as the grid spacing is decreased and the
extraction radius is increased.  The estimated errors in all the kick
results presented here due to the finite differencing grid spacing 
$h = M/0.8$ and extrapolation $r \rightarrow \infty$
are of the order of a few hundred $\mathrm{km \, s^{-1}}$.

\begin{table}
\begin{ruledtabular}
\begin{tabular}{cccrrrr}
  $P/M$ & $\phi_+$ & $L/M^2$
  & $V_\parallel$ & $V_\perp$ & $V$ & $E_\mathrm{rad} (\%)$\\
  \hline
    0.2379 &    0 & 1.064 & 5155.2 & -0.1 & 5155.2 & 4.9\\
    0.2665 &    0 & 1.192 & 6561.0 & -0.2 & 6561.0 & 7.6\\
    0.2739 &    0 & 1.225 & 6505.3 & -0.3 & 6505.3 & 8.4\\
    0.2851 &    0 & 1.275 & 5424.5 & -0.3 & 5424.5 & 10.1\\
    0.2952 &    0 & 1.320 & 3140.3 & -0.4 & 3140.3 & 11.8\\
  \hline
    0.2379 &   45 & 1.064 & 5483.9 & 709.7 & 5529.6 & 5.2\\
    0.2665 &   45 & 1.192 & 7614.9 & 932.4 & 7671.8 & 8.0\\
    0.2739 &   45 & 1.225 & 7830.2 & 930.5 & 7885.3 & 8.9\\
    0.2851 &   45 & 1.275 & 7227.6 & 778.5 & 7269.4 & 10.4\\
    0.2952 &   45 & 1.320 & 5026.4 & 384.7 & 5041.0 & 12.2\\
  \hline
    0.2379 &   90 & 1.064 & 4291.4 & 1074.6 & 4423.9 & 5.5\\
    0.2665 &   90 & 1.192 & 6485.7 & 1519.7 & 6661.3 & 8.7\\
    0.2739 &   90 & 1.225 & 6740.7 & 1528.0 & 6911.6 & 9.6\\
    0.2851 &   90 & 1.275 & 6001.4 & 1197.7 & 6119.7 & 11.5\\
    0.2952 &   90 & 1.320 & 3791.3 & 410.6 & 3813.3 & 13.1\\
\end{tabular}
\end{ruledtabular}
\caption{\label{tab:table1} 
  Configuration parameters $P/M$, $\phi_+$ and
  initial orbital angular momentum $L/M^2$ for the 
  cases with $\phi_- = 0^\circ$ and $S/M^2 = 0.2$. 
  The last four columns show the corresponding 
  kick velocities of the final \bh{} in $\KMS$ as 
  well as the \% of energy radiated.}
\end{table}

Table~\ref{tab:table1} gives the components of the recoil along the
initial orbital angular momentum $V_\parallel$ and in the initial
orbital plane $V_\perp$ in $\KMS$ (and the total recoil $V$) for the
cases with $\phi_- = 0^\circ$. Notice that the largest recoil
in this case happens
when $\phi_+ = 45^\circ$. Another important observation is that,
although the dominant component of the kick is $V_\parallel$, as we
increase $\phi_+$ a substantial component of the kick is also
generated in the orbital plane (see $V_\perp$).  The rightmost column
in Table~\ref{tab:table1} gives the $r \rightarrow \infty$
extrapolated radiated energy as a percentage of the total initial
energy of the binary. We see that large angular momentum increases the
energy radiated, up to very substantial values, as large as
$13\%$. For even larger angular momenta, we expect a falloff of the
radiated energy.

\begin{figure}
\begin{center}
\includegraphics[width=9.0cm,angle=0]{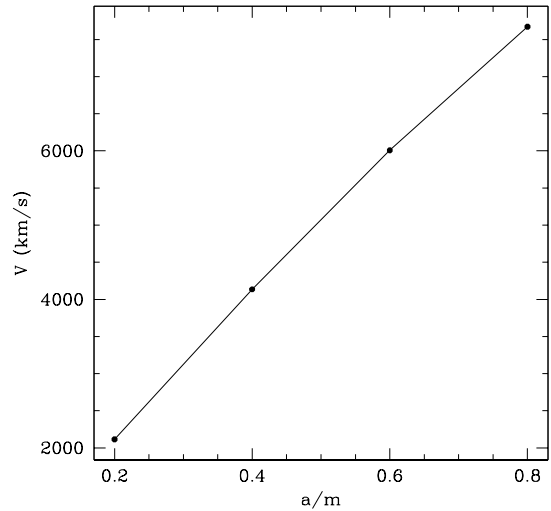}
\end{center}
\caption{Magnitude of the kick velocity as a function of the magnitude
  of the initial \bh{} spin ($a/m$) for the case ($P/M = 0.2665,
  \phi_+ = 45^\circ$)}
\label{fig:kicks_spin}
\end{figure}

For the case $P/M = 0.2665, \phi_+ = 45^\circ$ in
Table~\ref{tab:table1}, we have carried out simulations to investigate
the dependence of the kick with the initial \bh{} spins ($a/m$). The
results are displayed in Fig.~\ref{fig:kicks_spin}.  We find that, to
first order, the kick is proportional to $a/M_h$, as found in
quasi-circular orbits~\cite{2007ApJ...661..430H}. However, as the
initial spin of the \bh{s} grows, we found hints of the quadratic spin
dependence also obtained in quasi-circular
orbits~\cite{2007PhRvD..76l4002P}.

\begin{figure}
\begin{center}
\includegraphics[width=9.0cm,angle=0]{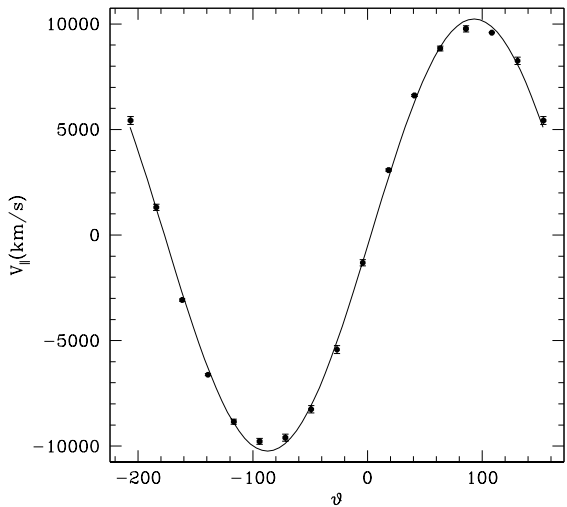}
\end{center}
\caption{Kick velocity $V_\parallel$ as a function of $\vartheta =
  \theta - \phi$ for the cases with spin orientation $\phi_+ = \phi_-
  = \phi$ and initial linear momentum $P/M = 0.3075$, corresponding to
  initial angular momentum $L/M^2 = 1.275$.  The solid line represents
  the fit $V_\parallel = -10,256\,\cos(\vartheta + 86^\circ)\,\KMS$.}
\label{fig:kick_phi}
\end{figure}

\begin{figure}
\begin{center}
\includegraphics[width=9.0cm,angle=0]{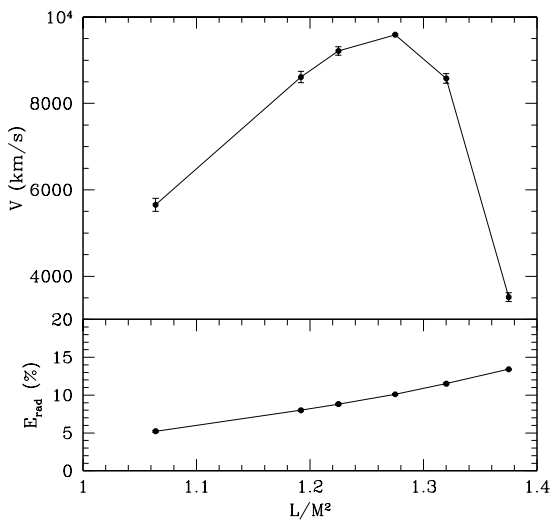}
\end{center}
\caption{Magnitude of the kick velocity (top panel) and \% of energy
  radiated (bottom panel) as a function of initial orbital angular
  momentum for spin orientation $\phi_+ = \phi_- = 45^\circ$.}
\label{fig:maxkick}
\end{figure}

\begin{figure}
\begin{center}
\includegraphics[width=9.0cm,angle=0]{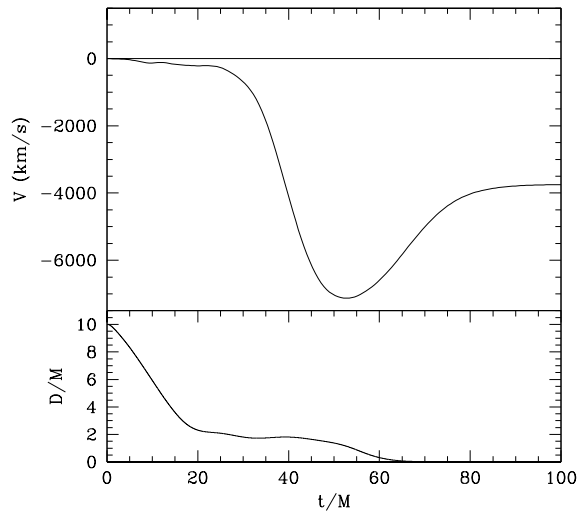}
\end{center}
\caption{Accumulation of the kick velocity extracted at $r=75\,M$ (top
  panel) and binary separation (bottom panel) as a function of time
  for the case $L/M^2 = 1.375, \phi_+ = \phi_+ = 45^\circ$ in
  Fig.~\ref{fig:maxkick}.} 
\label{fig:accum3} 
\end{figure}

The configuration we have found to yield the largest kick has the
spins anti-aligned $\phi_+ = \phi_- = \phi$, as in the case of
quasi-circular orbits~\cite{Gonzalez:2007hi,
  2007ApJ...659L...5C,2007PhRvL..98w1102C}, and initial linear
momentum $P/M = 0.3075$, corresponding to initial angular momentum
$L/M^2 = 1.275$. The kicks are essentially along the direction of the
initial orbital angular momentum. Figure~\ref{fig:kick_phi} shows the
kick velocity $V_\parallel$ as a function of $\vartheta = \theta -
\phi$, where $\vartheta$ measures the angle between the initial spin
and linear momentum vectors.  The solid line represents the fit
$V_\parallel = -10,239\,\cos(\vartheta + 86^\circ)\,\KMS$. This is the
same angular dependence found in quasi-circular
orbits~\cite{2007PhRvL..98w1102C}. Furthermore, as with circular
inspirals, the maximum kick is obtained for $\vartheta \approx
90^\circ$ although in these hyperbolic encounters the kick is
significantly larger, close to \kickmax.

To investigate the dependence on the initial orbital angular momentum,
we selected the case $\phi_+ = \phi_- = \phi = 45^\circ$ or $\vartheta
= 108^\circ$ in Fig.~\ref{fig:kick_phi}, which yields a kick magnitude
of $9,589\, \KMS$ and $15\,\%$ energy radiated, and carried out
simulations varying the initial momentum of the
\bh{s}. Figure~\ref{fig:maxkick} shows the magnitude of the kick
velocity $V$ (top panel) and the energy radiated $E_\mathrm{rad}$ in
\% of the initial energy (bottom panel) as a function of $L/M^2$. Note
that the maximum of $V$ does not occur for maximum
$E_\mathrm{rad}$. The case with the largest initial angular momentum,
$L/M^2 = 1.375$, has an interesting feature as displayed in the top
panel of Fig.~\ref{fig:accum3}. There is a pronounced \emph{anti-kick}
before the recoil reaches its final value.  The reason for this
anti-kick could be due to the fact that the plunge is not as
pronounced, appearing more \emph{circular-like}. That is, there is a
decrease in the rate at which the binary comes together, as one can
see in the bottom panel of Fig.~\ref{fig:accum3} during the time
interval $20-40\,M$. Thus, the flux vectors responsible for the kick
have more time to undergo the phase shift needed for the appearance of
an anti-kick~\cite{2008PhRvD..77d4031S}.

\begin{figure}
\begin{center}
\includegraphics[width=9.0cm,angle=0]{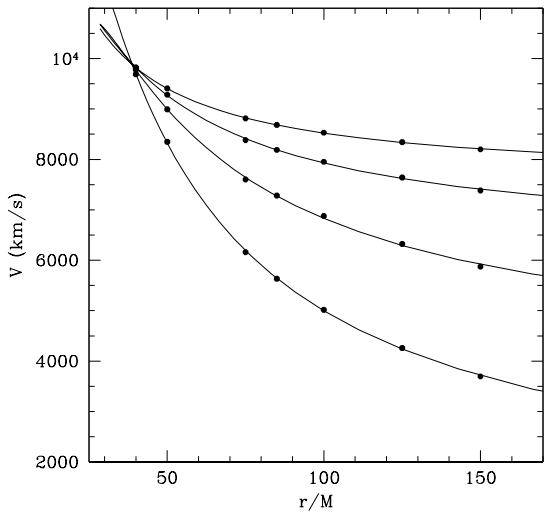}
\end{center}
\caption{Magnitude of the kick velocity as a function of the
  extraction radius for different resolutions in the case ($P/M =
  0.2665, \phi_+ = 45^\circ$). From bottom to top are respectively
  resolutions of $h = \lbrace M/0.4,
  M/0.5, M/0.63, M/0.8\rbrace$ at the extraction level. Dots are the
  data and lines correspond to the fit $V(r) = V_\infty^{(2)} +
  (K_1/r+ K_2/r^2)$.}
\label{fig:fit_2nd}
\end{figure}

In addition to the errors from extracting the gravitational recoil at
a finite radius, the values of the kicks are affected by numerical
finite differencing resolution.  To investigate these errors, we
selected the case $P/M = 0.2665, \phi_+ = 45^\circ$ from
Table~\ref{tab:table1} and carried out simulations with resolutions
$h = \lbrace M/0.4, M/0.5, M/0.63,
M/0.8\rbrace$ on the extraction level. Figure~\ref{fig:fit_2nd} shows
the data (points) and the corresponding fitting (lines) to $V(r) =
V_\infty^{(2)} + K_1/r+ K_2/r^2$. We find that at a given extraction
radius, the error in the kick scales as $h^3$, and the
kick itself grows with resolution.  Based on this $h^3$ behavior,
we estimate the error 
in the kick extrapolated as $r \to \infty$ computed using $h = M/0.8$
to be of the order of a few hundred $\mathrm{km \, s^{-1}}$, and we expect
all the kicks presented to be of a similar accuracy.

We have carried out a study of the gravitational recoil of the final
\bh{} in the merger of hyperbolic \bbh{} encounters. We have found
that in general the kick velocities for in-plane initial \bh{} spins
are significantly larger than those from the corresponding
quasi-circular mergers. Our results suggest that kicks as large as
\kickmax are possible.  We have also found that the dependence of the
kick on the initial magnitude of the \bh{'s} spins is similar to the
quasi-circular case. An analytic multipolar analysis of encounters for
similar configurations can be found in Ref.~\cite{Miller:2008en}. A
recent study by O'Leary, Kocsis and Loeb~\cite{2008arXiv0807.2638O}
has found that in dense population environments, there is a
non-negligible probability for close flybys or hyperbolic
encounters~\cite{Kocsis:2006hq}. Most of the cases they considered are
those in which after the first passage, the \bh{s} release enough
energy to become bound with large initial eccentricity. The hyperbolic
encounters we consider in the present work are the extreme case of
immediate merger.  Nonetheless, kicks with the magnitudes found in the
present study could lead to interesting astrophysical consequences.

Work supported in part by NSF grants PHY-0653443 (DS), PHY-0555436
(PL), PHY-0653303 (PL,DS), PHY-0114375 (CGWP), PHY-0354842 (RM) and
NASA grant NNG-04GL37G (RM). Computations carried out under LRAC
allocation MCA08X009 (PL,DS) and at the Texas Advanced Computation
Center, University of Texas at Austin. We thank M.~Ansorg, T.~Bode,
A.~Knapp and E.~Schnetter for contributions to our computational
infrastructure.


\end{document}